# Strain-induced perpendicular magnetic anisotropy in La$_2$CoMnO$_{6-\varepsilon}$ thin films and its dependence with film thickness

Regina Galceran[1,*], Laura López-Mir[1,*], Bernat Bozzo[1], José Cisneros-Fernández[1], José Santiso[2], Lluís Balcells[1], Carlos Frontera[1], and Benjamín Martínez[1]

[1]Institut de Ciència de Materials de Barcelona-CSIC, Campus UAB, E-08193 Bellaterra, Spain

[2]Institut Català de Nanociència i Nanotecnologia, ICN2-CSIC, Campus UAB, E-08193 Bellaterra, Spain

Abstract

Ferromagnetic insulating La$_2$CoMnO$_{6-\varepsilon}$ (LCMO) epitaxial thin films grown on top of SrTiO$_3$ (001) substrates present a strong magnetic anisotropy favoring the out of plane orientation of the magnetization with a large anisotropy field (~70 kOe for film thickness of about 15 nm). Diminishing oxygen off-stoichiometry of the film enhances the anisotropy. We attribute this to the concomitant shrinkage of the out of plane cell parameter and to the increasing of the tensile strain of the films. Consistently, LCMO films grown on (LaAlO$_3$)$_{0.3}$(Sr$_2$AlTaO$_6$)$_{0.7}$ and LaAlO$_3$ substrates (with a larger out-of-plane lattice parameter and compressive stress) display in-plane magnetic anisotropy. Thus, we link the strong magnetic anisotropy observed in LCMO to the film stress: tensile strain favors perpendicular anisotropy and compressive stress favors in plane anisotropy. We also report on the thickness dependence of the magnetic properties. Perpendicular anisotropy, saturation magnetization and Curie temperature are maintained over a large range of film thickness.

# I Introduction

Spintronic devices rely on the use of the spin degree of freedom of the electrons as control variable and are presently based on the generation and control of highly spin-polarized currents in ferromagnetic metals [1] or, more recently, on controlling pure spin currents [2]. In this context, ferromagnetic insulators may play an important role as spin sources or spin conductors [3]. In addition, they also have emerged as potential candidates for magnetically active barriers or spin filters [1]. Ferromagnetic insulating materials are scarce since in many cases ferromagnetic interactions are of exchange-type and driven by carriers. Among these rare materials, La$_2$CoMnO$_6$ and La$_2$NiMnO$_6$ double perovskites have been reported to be ferromagnetic insulators when cationic ordering (Co,Ni/Mn) is achieved [4,5]. Ferromagnetic ordering in these materials relies on the Goodenough-Kanamori-Anderson rules that predict a ferromagnetic interaction between Co$^{2+}$ or Ni$^{2+}$ ($t_{2g}^5 e_g^2$ and $t_{2g}^6 e_g^2$ respectively) and Mn$^{4+}$ ($t_{2g}^3$) cations when they are ordered in a fully alternating way in the lattice [6,7,8]. In addition, they have been intensively investigated recently because of claims of magnetodielectric response [9,10], which could be highly interesting for the implementation of new devices including tunable filters, magnetic sensors, and spin-charge transducers [11].

Much interest is added when the magnetization of these materials is found to be out-of-plane, as the possibility of controlling perpendicular magnetic anisotropy (PMA) opens the door to the implementation of high density magnetic memory devices [12,13]. Actually, in some cases, magnetic anisotropy appears to be sensitive to lattice distortion and point towards promising spintronic applications based on the ability to control magnetic properties through modification of strain conditions. It has been proved in several material systems that the easy magnetization axis can be changed from in plane to out of plane by means of tensile strain as in (GaMn)As [14], (IrMn)As, [15], (GaMn)(PN) [16], Fe garnets [17]; or by means of compressive strain in $La_{0.7}Sr_{0.3}MnO_3$ [18] and $SrRuO_3$ [19]. Other interesting systems where strain-induced anisotropy changes have been revealed are $Y_3Fe_5O_{12}$, which is widely used in microwave applications [20], CoFeB based systems [13,21,22] that additionally have functional properties such as giant tunnel magnetoresistance or $CoFe_2O_4$ and $CoCr_2O_4$ spinel systems that also present large magnetostriction [23,24,25,26,27]. Interestingly, strain tunable magnetocrystalline anisotropy has been found in double perovskite systems such as $Sr_2FeMoO_6$ [28] and $Sr_2CrReO_6$ [29,30].

In this paper, we investigate the effect of strain on magnetic anisotropy in $La_2CoMnO_{6-\varepsilon}$ (LCMO) thin films. We present a careful magnetic characterization of LCMO epitaxial thin films prepared by sputtering on top of different substrates that impose different structural strain. We also study the thickness dependence of the anisotropy on both compressive and tensile strain. Full cationic ordering of $Co^{2+}/Mn^{4+}$ in this double perovskite leads to a saturation magnetization of 6 $\mu_B$/f.u. and a Curie temperature, $T_C$, of about 225 K [4,31]. Both saturation magnetization and $T_C$ are very sensitive to cationic ordering and are substantially reduced in disordered samples [32,33]. Still our films show saturation magnetization values of about 6 $\mu_B$/f.u. irrespective to the substrate used and down to thickness of about 4 nm, thus indicating full cationic ordering. Remarkably, our results reveal strong perpendicular magnetic anisotropy in samples under tensile strain. We also investigate theoretically its possible origin, that we attribute to the spin-orbit coupling effect arising from $Co^{2+}$ ions in octahedral sites under tensile deformation. To the best of our knowledge PMA on LCMO has never been reported [10,32,33,34] and makes this material very appealing for implementation of magnetic tunnel junctions.

## II Experimental details

LCMO films have been prepared on top of (001) oriented $SrTiO_3$ (STO), $(LaAlO_3)_{0.3}(Sr_2AlTaO_6)_{0.7}$ (LSAT) and $LaAlO_3$ (LAO) substrates by means of RF magnetron sputtering. The target was prepared by solid state reaction according to the details published elsewhere [31]. Films have been grown at 900ºC (heater temperature) with 0.4 Torr of oxygen partial pressure, placing the substrate at 5 cm from the target, and with different *in situ* thermal treatments after deposition. Sample labels and preparation conditions are listed in Table I. Structural properties of thin films were studied by X-ray diffraction and reflectivity using a Rigaku Rotaflex RU200B diffractometer equipped with a rotatory anode, a Siemens D5000 diffractometer, a four-angle diffractometer with monochromatic Cu-K$_{\alpha 1}$ radiation (X´Pert MRD-Panalytical) and a Bruker D8 Advance GADDS system. Magnetization measurements were performed using a superconducting quantum interference device (SQUID, Quantum Design) as a function of temperature and/or magnetic field. For these measurements, we have applied the field perpendicular to the sample plane (out of plane, OP) or parallel to it (in-plane, IP). Magnetic torque measurements were carried out in a Physical Properties Measurement System (PPMS, Quantum Design) by using the torque-meter option. Surface topography of the films was investigated by atomic force microscopy (AFM), using an Asylum Research MFP-3D microscope in tapping mode, revealing a flat surface with terraces-and-steps morphology which follows the underlying STO surface morphology.

## III Results and Discussion

III.a *Magnetic anisotropy in LCMO//STO films.*

Figure 1 shows magnetization vs. temperature measured in sample A under an applied field of 1 kOe both OP and IP. The value obtained at low temperature with the field applied OP is nearly ten times larger than when field is applied IP. In addition, IP curve (FC branch) presents a small kink at about 100K. Similar kinks are also present for LCMO films grown by PLD, e.g. in Refs. [10,32,33,34],yet those works do not specify whether the measures were IP or OP. In our case, this anomaly is not present in the OP curve, even when its derivative is examined in detail. Thus, we discard that it comes from a second phase or from a second transition. This is reinforced by the fact that the ZFC branch of the IP $M(T)$ curve does not present any strange behavior at the same temperature.

In order to gain a deeper insight into these features, we have measured the magnetization curves $M(H)$ (at $T$= 10 K and the magnetic field applied both IP and OP) for films B, C, and D. Preparation conditions of these three samples differ in the annealing conditions after preparation. For that reason samples present different (and increasing) oxygen contents as reported in Ref. [35]. Results corresponding to these three samples are shown in Fig. 2. It is evident from the figure that when the field is applied OP the magnetization is higher than when it is applied IP. On the other hand, the difference between the two directions is enhanced for films with larger oxygen content. Hence, these results show that, for the three samples, the easy magnetization axis is OP, i.e. the films exhibits perpendicular magnetic anisotropy (PMA). However, the strength of the anisotropy changes with oxygen content: the larger the oxygen content the stronger the anisotropy. In addition, saturation magnetization also increases with oxygen content, thus suggesting that oxygen deficiency promotes cationic disorder.

To determine the anisotropy field of samples B and D, we have performed magnetic torque measurements by rotating the sample with respect to the magnetic field directions moving from OP ($\psi= 0º$, 180º) to IP ($\psi= 90º$, 270º). At every angle, we measured the torque that the field makes on the sample due to its magnetic moment. Results are plotted in Fig. 3 for three different temperatures and exhibit the typical shape corresponding to a system with uniaxial anisotropy [36]. These measurements corroborate that the easy magnetization axis is perpendicular to the film plane and that the magnetic anisotropy is stronger in the highly oxygenated sample. The anisotropy field can also be inferred from the maximum value of the torque, and the saturation magnetization through the expressions:

$$k_1 = {\tau_M}/{V}$$
$$H_A = {2k_1}/{M_S}$$

Being $\tau_M$ the maximum torque, $V$ the sample volume, $k_1$ the effective anisotropy constant and $M_S$ the saturation magnetization [36]. This renders $H_A$= 67.1 and 27.5 kOe for samples D and B, respectively.

This strong anisotropy explains the large difference between ZFC and FC $M(T)$ in Fig. 1 which must be ascribed to the fact that the magnetic field used (0.1 T) is considerably smaller than the coercive field and the anisotropy field: at the end of the zero field cool process, a similar number of domains with magnetization up and magnetization down must be expected, rendering a small value of the overall magnetization. The small field applied is not able to switch the domains up to a temperature near $T_C$. Concerning the kink in the IP-$M(T)$ curve (Fig. 1), the fact that it is not present neither in ZFC branch nor in the OP curve suggests that the anomaly could appear due to a competition between anisotropy and cooling field.

As we have reported in a previous work [35], the change of the oxygen stoichiometry promotes a reorientation of the LCMO lattice on top of the STO substrate. Low oxygen content makes the *c*-axis of LCMO to lay IP while large oxygen content turns *c*-axis OP. Nevertheless, easy magnetization axis is found to be OP independently of the *c*-axis orientation (sample B has *c*-axis IP, and sample D has *c*-axis OP [35] but both present OP anisotropy). This evidences that crystallographic orientation does not determine the easy magnetization axis. Hence, the change in the oxygen content is reflected in a change in the strength of the anisotropy but not in its direction. In this regard, it is worth mentioning here that in many perovskites the orientation of the crystallographic cell is entirely determined by how metal-oxygen octhaedra rotate. Octahedra rotation in bulk LCMO is of type $a^-a^-c^+$ according to Glazier's notation. On another hand, films B, C, and D are in-plane fully strained by the substrate. This means that LCMO unit cell is clamped to the substrate cell. Therefore, the reorientation of *c*-axis only implies a change of the direction where successive octahedra rotate in the same sense ($c^+$) from parallel to perpendicular to the film. Hence, the reorientation of the crystallographic cell, by itself, does not imply a change of the arrangement of Mn and Co cations nor a change of the Mn-Co distances (only La and O are affected). In fact, lattice strain fixes Mn-Co distances.

We have also observed that oxygen content shortens the OP lattice parameter: sample B has a larger OP parameter than sample D [35]. This shortening is a consequence of the lattice volume shrinkage as oxygen approaches the nominal stoichiometric value (oxygen vacancies promote the appearance of $Mn^{3+}$ ions with ionic radius larger than $Mn^{4+}$ ions). This volume shrinkage also produces an increase of the lattice mismatch as lattice is under tensile strain (lattice mismatch would be +0.6% for full oxygen stoichiometry). On the other hand, the shortening of OP lattice parameter implies shorter Mn-Co distances in the perpendicular direction. Therefore, the change of the anisotropy could be attributed to the change in the out of plane lattice parameter, with its concomitant reduction of cationic distances in the perpendicular direction, which would promote a reinforcement of orbital bonds in this direction. In addition, lattice strain can be a source of anisotropy in systems showing magnetostriction [37,38]. To the best of our knowledge magnetostriction of LCMO has not been reported but huge anisotropic one has been found in other cobalt-based perovskites (e.g. $La_{1-x}Sr_xCoO_3$ [39]).

To clarify the role of structural strain on the magnetic anisotropy we have grown LCMO films on substrates with different in-plane lattice parameter (LSAT and LAO, with lattice mismatch of -0.3% and -2.3% respectively) under the same conditions used for growing on top of STO substrates.

III.b *Magnetic anisotropy in LCMO//LSAT and LCMO//LAO films.*

The structural features of LCMO films grown on LSAT and LAO substrates have been studied by X-ray diffraction. Reciprocal space maps around (103) substrate peaks (Figs. 4a and 4b) show that sample E (LCMO//LSAT) is in-plane fully strained (IP lattice parameter is 3.87 Å) while sample F (LCMO/LAO) is partially relaxed (IP parameter estimated from the reciprocal space map is about 3.84 Å). High resolution θ/2θ scans around (002) substrate peak have been analyzed by using the expressions given in Ref. [40] (observed and calculated intensities, and the difference between both, are plotted in Figs. 4c and 4d). Fitting of the data allows estimating OP lattice parameters that are about 3.906(3) Å and 3.912(3) Å for samples E and F respectively. These values are, in both cases, larger than their respective IP parameters and larger than *c* parameter obtained for sample B [3.901(3) Å].

Figure 5 shows the temperature dependence of the magnetization measured under a field of 1 kOe applied perpendicular (OP) and parallel (IP) to sample plane for samples E and F. In both cases, the Curie temperature is around 230 K indicating an optimum oxygen content and film quality. It is also evident that the magnetization IP reaches values much larger than OP, implying an IP orientation of the easy magnetization axis. This is further confirmed by *M*(*H*) curves (Fig. 6) where the magnetization measured IP is always larger than that measured OP. In addition, both *M*(*T*) and *M*(*H*) curves suggest that the IP anisotropy is larger in LCMO//LAO than in LCMO//LSAT. This would be in accordance with the larger OP parameter of the former.

On the other hand, Fig. 6 shows that saturation magnetization reaches $6\mu_B$/f.u., the expected value for films displaying a good Co/Mn cationic ordering. Coercive fields for IP $M(H)$ curves are around 7 and 8 kOe for samples E and F respectively, while OP ones are of 450 and 600 Oe. Values of the coercive field for the hysteresis loops measured with $H$ along the hard magnetization axis are much smaller than in the case of LCMO//STO (as can be observed in Fig. 2c the coercive fields for easy and hard axis are nearly the same, e.g. ~7 kOe for sample D).

We have also examined how the anisotropy varies with film thickness. In thin films the effective anisotropy constant can be expressed as the sum of two terms, $k_1 = k_v + 2k_s/t$ [37,41], where $k_v$ is the volume, $k_s$ the surface term and $t$ the film thickness. The surface term takes into account the anisotropy appearing due to the interface and its intrinsic translational symmetry breaking [42]. This term becomes more relevant for thinner films, and usually favors OP orientation of the magnetic moment. The two samples of smaller thickness grown on LSAT and LAO (samples G and H respectively) show the prevalence of the IP orientation of the easy magnetization axis. This can be observed in M(H) curves (Fig. 7). Even though magnetization IP reaches larger values than OP, the difference between both is not as relevant as in the case of thicker samples, the differences in the coercive field and remanent magnetization reinforce that the easy axis lays IP. This result shows that $k_s$ is positive (thus favoring OP magnetization) and $k_v$ is negative (therefore favoring IP magnetization), being $k_v$ dominant in all the cases studied.

III.c *Thickness dependence of magnetic properties of LCMO//STO films.*

As mentioned above, it is interesting to characterize the behavior of LCMO films as a function of thickness for two reasons. First, to examine the properties of very thin films (~4 nm) that could be useful as active insulating barriers; and second to study whether or not the PMA found is present in thicker films.

The strain state as a function of the thickness of the film has been studied by X-ray diffraction. θ/2θ scans around the (002)-STO substrate peak show that OP lattice parameter is nearly constant with thickness: it runs from 3.868(8) Å for t= 4 nm to 3.877(2) Å for t= 66 nm. Moreover, a reciprocal space map around (103) reveals that sample L is fully strained by the substrate (not shown). Thus, we conclude that no lattice relaxation occurs in the range studied.

Figure 8(a) shows the temperature dependence of the magnetization measured with the magnetic field applied OP for LCMO//STO films of different thicknesses. It can be appreciated that the shape of the magnetization curves is very smooth and that the Curie temperature hardly depends on $t$ (for $t \geq 8$ nm). For the thinnest studied sample ($t=4$ nm) $T_C$ is shifted down to 200 K. The shape of the OP $M(T)$ curve is indicative of PMA for all values of $t$. This is further confirmed by $M(H)$ hysteresis loops (measured also OP) plotted in Fig. 8(b). All samples present a remarkable square shape with a square ratio (remanence/saturation) very close to 1, proving the strong magnetic anisotropy of the films and that the easy axis is OP. Coercive field remarkably grows when reducing film thickness ($H_C$= 1.5 and 0.55 T for $t$= 4 and 66 nm films respectively). Loops present a small jump near $H$=0, in an amount that is nearly independent of $t$ (at least for $t$< 34 nm), which could indicate the presence of some region at the interface or at the surface with a different coercive field. This is so because interfaces may introduce disorder and changes on the strength of magnetic interactions and anisotropy because of translational symmetry breaking and border effects. Those effects could be especially relevant when the material is strongly anisotropic as in the present case.

In order to gain insight into the PMA found, we have performed torque measurements in sample L, the thickest one, which render a $k_1$ value smaller than for sample D (values are 1.64 *vs* 1.16 MJ/m$^3$ for samples D, $t$= 15 nm, and L, $t$= 66 nm, respectively). This reduction can be ascribed to the smaller contribution of $k_s$ term in the thicker sample. The values of $k_1$ found indicate that $k_v \approx$ 1.02 MJ/m$^3$ and $k_s \approx$ 9.18 mJ/m$^2$. The volumetric term $k_v$ takes into account the magnetocrystalline, strain, and shape anisotropy terms. The latter term always tends to place the magnetization in plane and never out of plane. Magnetocrystalline and strain anisotropy terms can either be IP or OP. Thus, the change from OP to IP anisotropy in LCMO//LSAT and LCMO//LAO films must be due to a weakening of the OP strain anisotropy, or even a change in its sign, making the shape anisotropy to become dominant. In both cases the surface term is found to be OP.

III.d *Origin of PMA and its dependence on film strain*

As mentioned, different magnetic epitaxial films show a change in the anisotropy direction under different strain [14-27]. Of special relevance is the dependence on strain of the magnetic anisotropy found in Co$^{2+}$-containing spinels. In CoCr$_2$O$_4$ a compressive strain favors PMA, while a tensile strain favors in plane anisotropy [23]. On the contrary, CoFe$_2$O$_4$ coincides with the behavior of LCMO reported here in the sense that a tensile stress induces PMA [24,25], while a compressive strain induces in-plane anisotropy [26,27]. In both systems, magnetic anisotropy is attributed to Co$^{2+}$ ions [23]. CoCr$_2$O$_4$ is a normal spinel, where Cr ions occupy octahedral sites and Co ones occupy tetrahedral sites of the structure. In contrast, CoFe$_2$O$_4$ is an inverse spinel, in which tetrahedral sites are occupied by Fe and octahedral ones by Fe and Co. According to theoretical calculations, compressive strain must favor PMA for Co$^{2+}$ in tetrahedral environment (CoCr$_2$O$_4$) and an in plane easy magnetization axis when it is placed in octahedral environment (CoFe$_2$O$_4$) [23]. Consistently, in LCMO, where Co$^{2+}$ is in an octahedral environment, we found that a compressive strain favors an in plane easy axis.

Theoretical calculations in Ref. [23] for iron spinel consider the spin-orbit coupling in Co$^{2+}$ and the effect of the crystal field departure from a perfect octahedral environment, as perturbations of the other terms of the Hamiltonian. In Appendix A, we present a theoretical calculation on how, departing from a perfect octahedron, tetragonal strain (tensile or compressive) must affect the magnetic anisotropy of Co$^{2+}$. Although the case in real LCMO is not exactly the same, this result helps to understand our findings as it fairly predicts PMA for tensile strain and an in-plane easy axis for compressive strain.

## IV Summary and conclusions

We have shown that LCMO films grown on top of STO have strong perpendicular magnetic anisotropy. Torque measurements reveal that the anisotropy field shows a close correlation with the degree of oxygen content of the films: the larger the oxygen content, the larger the anisotropy field. We have related this fact to the change in the out of plane cell parameter of LCMO with the oxygen content (the larger the oxygen content, the larger the anisotropy and the shorter the out of plane cell parameter) rather than to the change in the crystallographic cell orientation. In a context of tensile mismatch between substrate and film, the shrinkage of the out of plane lattice parameter implies a larger strain. We have further investigated the effect of the strain by growing LCMO on top of LSAT and LAO with smaller in plane parameter. This gives rise to films with larger out of plane parameter and a compressive instead of tensile strain. As a result the easy magnetization axis goes from OP to IP. This proves that strain rules the magnetic anisotropy of these films. We have investigated the possible origin of this phenomenon, in accordance with previous literature [23,43], by considering an idealized case in which strain mainly introduces a tetragonal distortion in otherwise cubic crystal field. Such scenario predicts our results: PMA for tensile strain and IP easy axis for compressive strain.

We have also studied the dependence of the magnetic properties and anisotropy on the film thickness. We have found no relaxation of the lattice and ferromagnetism for all the thicknesses studied (between 4 and 66 nm). The Curie temperature is only reduced for very thin films: for samples of about 8nm and above $T_C$ is ~230K, but for 4 nm film it decreases down to ~200 K. In all cases, films grown on STO present PMA and the anisotropy constant decreases with thickness. The obtained values evidence that, besides the contribution to the anisotropy from the interface term, the volume term also contributes positively to the PMA. The existence of the interface term is reinforced by the thickness dependence of the magnetic anisotropy for films grown on top of LSAT and LAO substrates.

## Acknowledgements

Authors thank Dr. X. Martí for providing the code for calculating the diffracted intensity according to expressions in Ref. [40]. We acknowledge financial support from the Spanish Ministry of Economy and Competitiveness through the "Severo Ochoa" Programme for Centres of Excellence in R&D (SEV-2015-0496), and projects MAT2012-33207 and MAT2015-71664-R.

## Appendix A: Effect of tetrahedral crystal field and spin-orbit coupling on $Co^{2+}$ in octahedral environment

The objective of these theoretical calculations is to find the magnetic anisotropy expected for the ideal case in which $Co^{2+}$ ion is placed in an octahedron with tetrahedral symmetry (compressed/expanded along one of the main axes). We start from the case where the octahedron has a perfect cubic symmetry and introduce the tetrahedral distortion and spin orbit coupling (SOC) as a perturbation. For this, we follow a procedure similar to that presented in Refs. [43,44] for $CoCl_2$ in which the octahedron departs from cubic symmetry due to a trigonal distortion (it is compressed/expanded along one of the main diagonals of the cube).

The ground state of a free $Co^{2+}$ corresponds to a $^4F$ term (L=3, S=3/2), that under the effect of a crystal field produced by an octahedral environment with cubic symmetry splits in three levels, 2 triplets and one singlet. The lowest level is the triplet $^4T_1$ whose eigenstates are [45]:

$$\varphi_0 = |3\,0\rangle$$

$$\varphi_+ = \sqrt{3/8}\,|3\,\text{-}1\rangle + \sqrt{5/8}\,|3\,3\rangle$$

$$\varphi_- = \sqrt{3/8}\,|3\,1\rangle + \sqrt{5/8}\,|3\,\text{-}3\rangle$$

The tetrahedral distortion of the octahedron (crystal field) and the SOC interaction are introduced as a perturbation $H' = H_{LS} + H_{CF}$ in the Hamiltonian. According to first order perturbation theory, one must find the diagonalization of $H'$ matrix in the ground state of unperturbed Hamiltonian. $H_{LS}$ is expressed as $k\lambda \vec{L}\cdot\vec{S}$, where $\lambda$ is the spin-orbit constant, that is expected to be negative (for *d*-shells with more than half filling), and $k$ is the ''orbital reduction factor'' ($k \lesssim 1$) [43,44]. $H_{CF}$ is the deviation of crystal field from cubic symmetry. According to Ref. [45] tetragonal crystal field is expressed as:

$H_{CF}^{tet} = A_2^0 r^2 Y_2^0 + A_4^0 r^4 Y_4^0 + r^4\left(A_4^4 Y_4^4 + (A_4^4)^* Y_4^{-4}\right)$ while for a cubic symmetry it reduces to $H_{CF}^{cub} = A_4^0 r^4 \left[Y_4^0 + \left(\frac{5}{14}\right)^{1/2}\left(Y_4^4 + Y_4^{-4}\right)\right]$. Thus, the tetragonal field differs from cubic one in a term on $Y_2^0$ and on the fact that of $A_4^0$ and $A_4^4$ are no longer related. We have, in a first approximation, ignored this second fact and considered only the first one. Using this approximation, the $H_{ij}^{CF} = \langle\varphi_i|H_{CF}|\varphi_j\rangle$ matrix elements different from zero are: $H_{00}^{CF} = 2\epsilon_{CF}$ and $H_{++}^{CF} = H_{--}^{CF} = -\epsilon_{CF}$. This means that (before considering SOC), $\varphi_0$ state, or equivalently $|3\,0\rangle$, that is mainly oriented along z axis, is more affected by the tetragonal distortion of the crystal field than $\varphi_{\pm 1}$. In the case of tensile strain, where basal distances of the octahedra are larger than apical ones, $\epsilon_{CF}$ is positive and $|3\,0\rangle$ is the state with higher energy while the other two sates of $^4T_1$ triplet have lower energy. This is in agreement with the expected degeneration of the ground state under tensile stress [23]. In the case of compressive strain $\epsilon_{CF}$ is negative, and $|3\,0\rangle$ becomes the ground state (crystal field only).

To consider the spin orbit interaction, $H_{ij}^{LS}$ matrix elements must also be calculated by using:

$\vec{L}\cdot\vec{S} = L_z S_z + \frac{1}{2}(L_+ S_- + L_- S_+)$; $L_\pm |LM\rangle = \sqrt{(L \pm M + 1)(L \mp M)}|L \pm 1\rangle$ (and the equivalent for $S_\pm$ operators).

The three levels of $^4T_1$ term must be combined with the four possible spin states (S=3/2) thus giving rise to twelve states. The lowest energy level is a Kramer's doublet that corresponds to (assuming λ<0):

$\psi_- = \alpha|\varphi_- -\frac{3}{2}\rangle + \beta|\varphi_0 -\frac{1}{2}\rangle + \gamma|\varphi_+ \frac{1}{2}\rangle$; $\psi_+ = \alpha|\varphi_+ \frac{3}{2}\rangle + \beta|\varphi_0 \frac{1}{2}\rangle + \gamma|\varphi_- -\frac{1}{2}\rangle$

where α, β, and γ are coefficients that only depend on $a = \frac{2\epsilon_{CF}}{-3k\lambda}$.

To characterize the anisotropy, we calculate the gyromagnetic factor for a field applied along the 4-fold axis of the distorted octahedra ($g_z$) and for a field applied in a perpendicular direction ($g_x$). Gyromagnetic factors are calculated through the energy splitting (ΔE) of the Kramer's doublet when applying a magnetic field along the specified direction, and are found by the diagonalization (within the Kramer's doublet) of the Hamiltonian:

$$H_Z = \mu_B(k\vec{L} + 2\vec{S})\cdot \vec{H}$$

And applying $g=\Delta E/\mu_B H$ for each direction. This renders:

$\frac{g_z}{g_x} = \frac{(3+\frac{3}{2}k)\alpha^2+\beta^2-(1+\frac{3}{2}k)\gamma^2}{2\alpha\gamma\sqrt{3}+k\beta\gamma\frac{3}{2}\sqrt{2}+2\beta^2}$

Figure 9 shows the plot of the ratio of these gyromagnetic factors in front of $a = \frac{\epsilon_{CF}}{-2k\lambda}$ for $k$=0.9 calculated after finding of α, β, and γ coefficients numerically. This ratio crosses the value $g_z/g_x$=1 when $a$ (and then $\epsilon_{CF}$) changes its sign. So under tensile strain, with positive values of $\epsilon_{CF}$, we find that $g_z > g_x$ and thus PMA is predicted, while for compressive strain, with negative values of $\epsilon_{CF}$, $g_x > g_z$ and an in-plane anisotropy must be expected.

Table I: Thickness and annealing conditions of the $La_2CoMnO_6$ films prepared by RF magnetron sputtering. All samples were grown at a partial oxygen pressure of 0.4 Torr and annealed at 900ºC.

| Name | t (nm) | Substrate | Annealing $p_{O2}$ (Torr) | Annealing time (h) | Cooling rate (ºC/min) |
|---|---|---|---|---|---|
| A | 15 | STO | 400 | 2 | 10 |
| B | 15 | STO | $2\times10^{-6}$ | 1 | 10 |
| C | 15 | STO | 400 | 1 | 10 |
| D | 15 | STO | 400 | 1 | 1 |
| E | 15 | LSAT | 400 | 2 | 10 |
| F | 15 | LAO | 400 | 2 | 10 |
| G | 4 | LSAT | 400 | 2 | 10 |
| H | 4 | LAO | 400 | 2 | 10 |
| I | 4 | STO | 400 | 2 | 10 |
| J | 8 | STO | 400 | 2 | 10 |
| K | 34 | STO | 400 | 2 | 10 |
| L | 66 | STO | 400 | 3 | 10 |

FIGURES

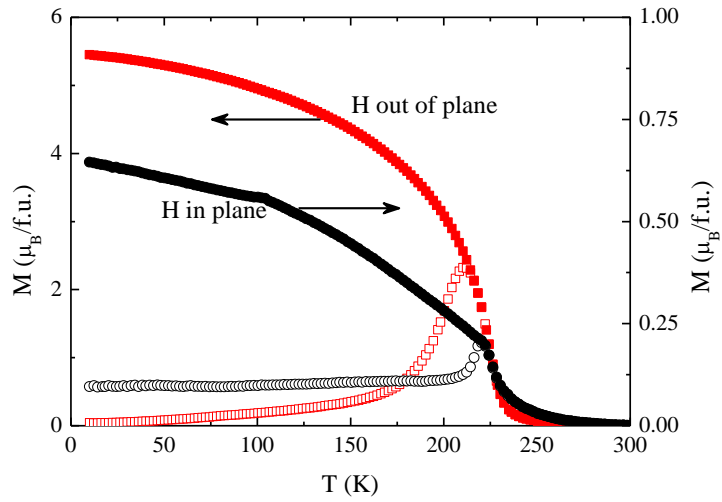

Figure 1. Magnetization versus temperature measured after ZFC (open symbols) and FC (solid symbols) for sample A (t=15 nm) with magnetic field of 1 kOe applied out of plane (red squares, left axis) and in plane (black circles, right axis).

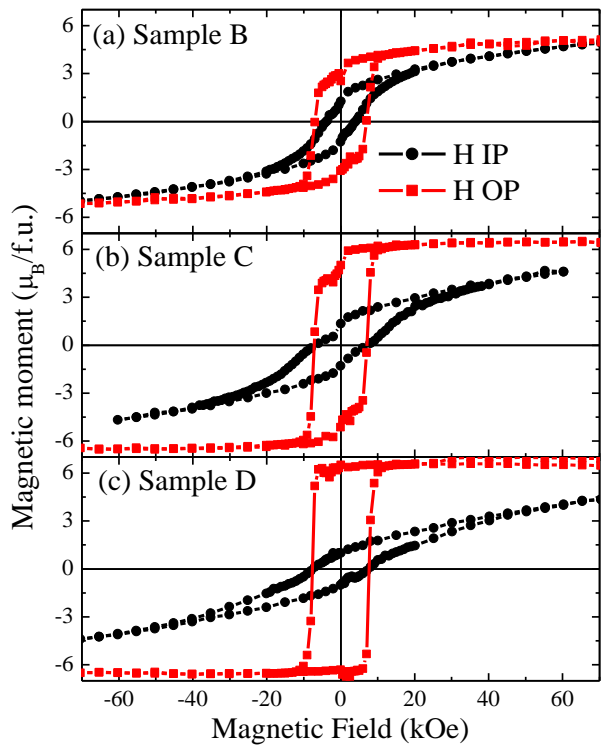

Figure 2. Magnetization hysteresis loops, at T=10K, with magnetic field applied out of plane (red squares) and in plane (black circles) for samples prepared under different annealing

conditions. Thickness of these samples is around 15 nm.

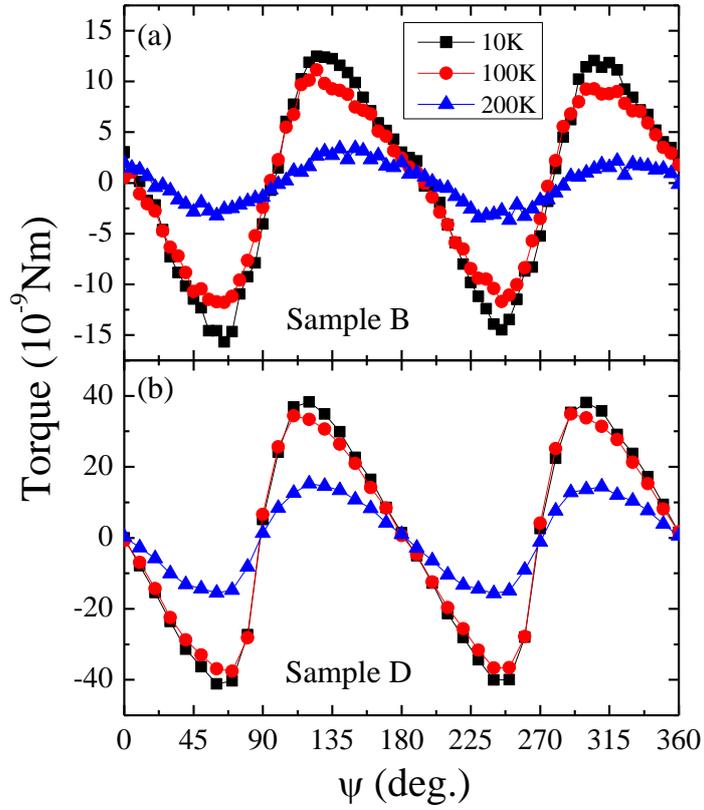

Figure 3. Torque needed to rotate samples B (a) and D (b) in a magnetic field of $\mu_0 H$=8 T at different temperatures. Thickness of these samples is around 15 nm.

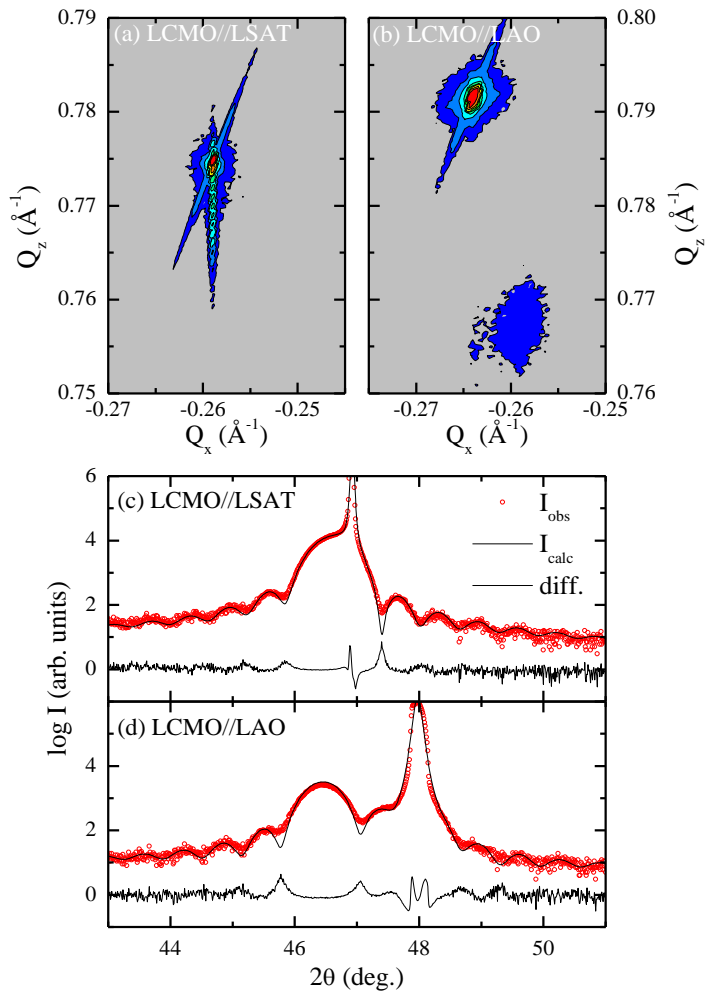

Figure 4. Reciprocal space maps collected around (-103) diffraction peak of LCMO grown on (a) LSAT (sample E) and (b) LAO (sample F). Panels (c) and (d) show the refinement of the high resolution θ/2θ scans collected for the same samples respectively. Thickness of these samples is around 15 nm.

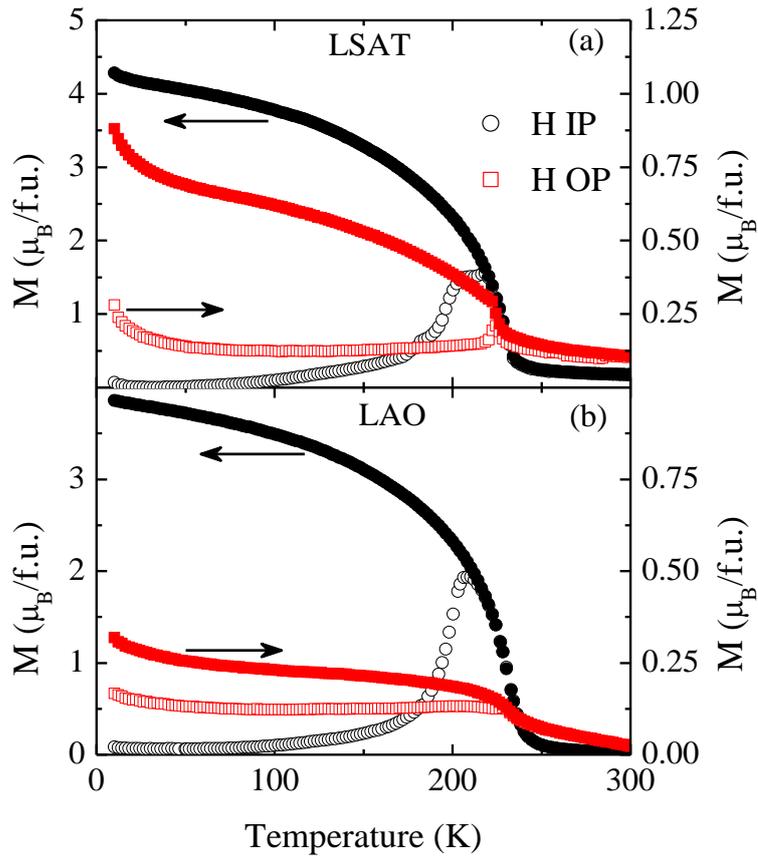

Figure 5. Magnetization versus temperature measured under a field of 1kOe after ZFC (open symbols) and FC (solid symbols) for samples (a) E (over LSAT) and (b) F (over LAO) with magnetic field in plane (black circles, left axes) and applied out of plane (red squares, right

axes). Thickness of these samples is around 15 nm.

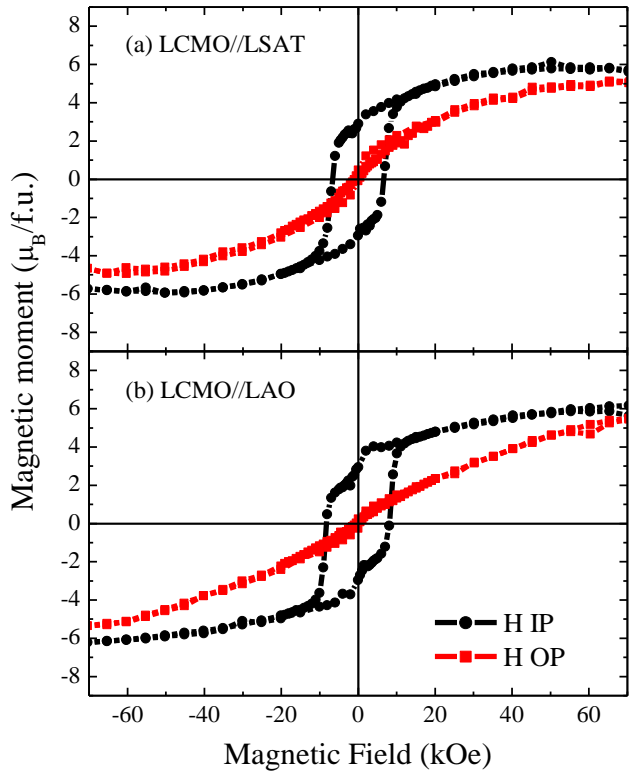

Figure 6. Magnetization versus field measured at T= 10K, for samples (a) E (over LSAT) and (b) F (over LAO) with magnetic field in plane (black squares) and applied out of plane (red squares). Thickness of these samples is around 15 nm.

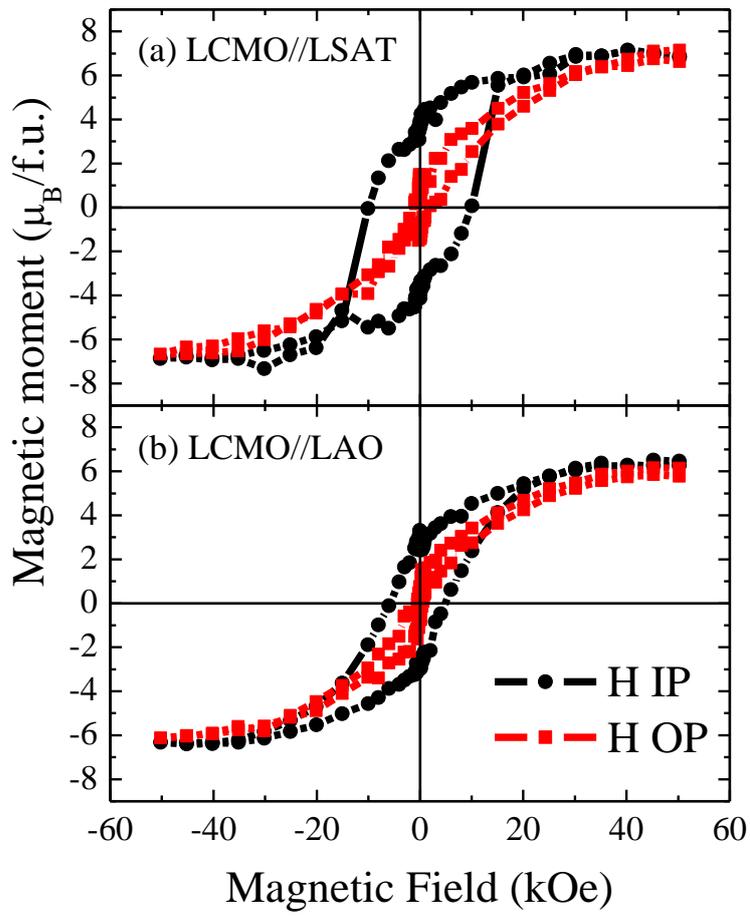

Figure 7. Magnetization versus field measured at T= 10K, for samples (a) G (over LSAT) and (b) H (over LAO) with magnetic field in plane (black squares) and applied out of plane (red squares). Thickness of these samples is around 4 nm.

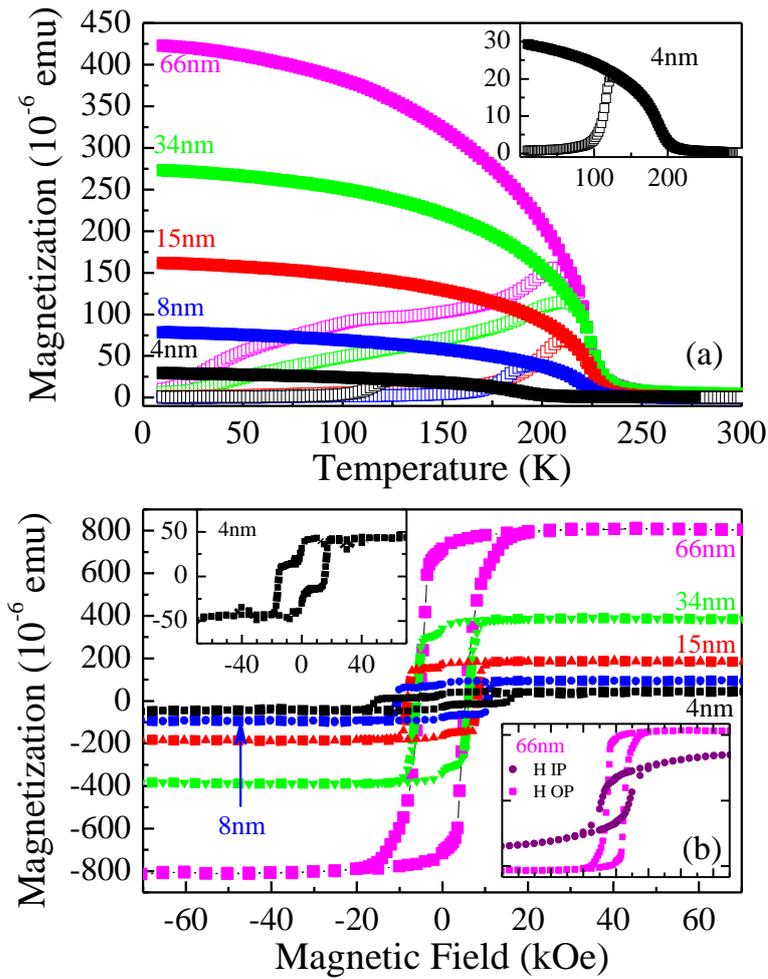

Figure 8. Thickness dependence of the magnetic properties of LCMO//STO. (a) OP magnetization under 1kOe after ZFC (open symbols) and FC (solid symbols) for different film thickness; the inset shows in detail that of 4 nm sample. (b) OP hysteresis loops measured at 10K with field OP; top left inset shows in detail the loop measured for 4nm thick sample; right bottom inset shows the loops obtained for 66nm thick sample with field OP and IP.

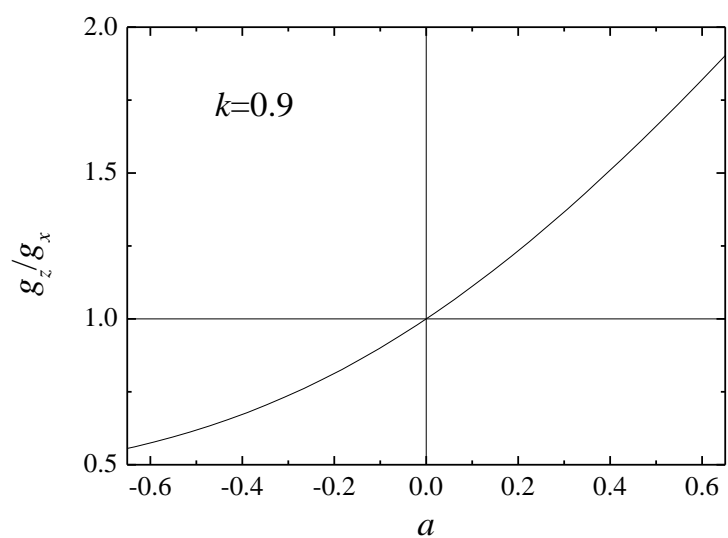

Figure 9 Coefficient of gyromagnetic factors in perpendicular and parallel directions as a function of the ratio between tetrahedral crystal field and spin orbit coupling $\left(a \equiv \frac{2\epsilon_{CF}}{-3k\lambda}\right)$.